 \documentclass[final,5p,times,twocolumn]{elsarticle}

 \usepackage{graphics}
 \usepackage{graphicx}
\usepackage{epsfig}
\usepackage{subfigure}

\usepackage{amssymb}

\usepackage[hyperindex,breaklinks]{hyperref} 

\usepackage{mathtools}
\usepackage{amsmath}
\usepackage{xspace}
\usepackage{atlasphysics}

\newcommand{\Zg}{\Zboson/\gamma^*}

\usepackage{preprintcover}  

\PreprintCoverPaperTitle{Measurement of the top quark pair production cross section with ATLAS in $pp$ collisions at $\sqrt{s}=7$ TeV using final states with an electron or a muon and a hadronically decaying $\tau$ lepton}  

\PreprintIdNumber{CERN-PH-EP-2012-102}  

\PreprintCoverAbstract{
A measurement of the cross section  of 
top quark pair production in proton$-$proton collisions  
recorded with the ATLAS detector at the LHC at a centre-of-mass energy of 7 TeV 
 is reported.
The data sample used corresponds to
an integrated luminosity of 2.05~${\rm fb}^{-1}$.
Events with an isolated electron or muon
and a $\tau$ lepton decaying hadronically are used. 
In addition, a large missing transverse momentum and two or more energetic jets 
are required. At least one of the jets
must be identified as  
originating from a $b$ quark. The measured cross section, 
$
\sigma_{t\overline{t}} =  186\pm 13  \stat  \pm 20\syst \pm 7  (\text{lumi.})  ~\rm{pb}
$,
  is in good agreement with the Standard Model prediction.
}  

\PreprintJournalName{Physics Letters B}  

\begin{document}

\begin{frontmatter}



\title{Measurement of the top quark pair cross section with ATLAS in $pp$ collisions \\ at $\sqrt{s}=7$ TeV
using final states with an electron or a muon \\ and a hadronically decaying $\tau$ lepton}


\author{The ATLAS Collaboration}

\address{}

\begin{abstract}
A measurement of the cross section  of 
top quark pair production in proton$-$proton collisions  
recorded with the ATLAS detector at the LHC at a centre-of-mass energy of 7 TeV 
 is reported.
The data sample used corresponds to
an integrated luminosity of 2.05~${\rm fb}^{-1}$.
Events with an isolated electron or muon
and a $\tau$ lepton decaying hadronically are used. 
In addition, a large missing transverse momentum and two or more energetic jets 
are required. At least one of the jets
must be identified as  
originating from a $b$ quark. The measured cross section, 
$
\sigma_{t\overline{t}} =  186\pm 13  \stat  \pm 20\syst \pm 7  (\text{lumi.})  ~\rm{pb}
$,
  is in good agreement with the Standard Model prediction.
\end{abstract}


\end{frontmatter}

\bibliographystyle{model1-num-names}
\bibliography{<your-bib-database>}

\section{Introduction}
\label{sec:Introduction}

Measuring the top quark pair ($t \overline{t}$) production cross section
($\sigma_ {t\overline{t}}$) in different decay channels is of interest
because it can open a window to physics beyond the Standard Model
(SM). In the SM, the top quark decays with a branching ratio close to 100\% 
 into a $W$ boson and a $b$ quark, and $t \overline{t}$ pairs are
identified by either the hadronic  or leptonic decays of the $W$ bosons
and the presence of additional jets. 
ATLAS has previously used  the single-lepton channel
~\cite{ATLAS_ljets}, and the dilepton channels
including only electrons and muons~\cite{ATLAS700_dil} to
perform cross-section measurements.
 
The  large cross section for $t \overline{t}$
production at the LHC provides an opportunity to measure $\sigma_ {t\overline{t}}$ using final states with 
an electron or a muon and a $\tau$ lepton
with high precision. The $\sigma_ {t\overline{t}}$ in this channel has been measured at the
Tevatron with 25\% precision \cite{D0} and recently by the CMS Collaboration at the LHC 
with 18\% precision \cite{CMSltau}. A deviation from $\sigma_ {t\overline{t}}$ measured in other channels would be an
indication of non-Standard Model decays of the top quark, such as a decay to a charged Higgs ($H^+$) and a $b$ quark 
with $H^+$ decaying to a $\tau$ lepton and a $\tau$ neutrino, or contributions from non-Standard Model processes~\cite{MSSM,Han,Yue}. ATLAS has set upper limits on the branching ratio of top quark decays to an $H^+$ bosons decaying 
to a $\tau$ lepton and a neutrino \cite{AtlasH+}.

This analysis uses 2.05~fb$^{-1}$ of data collected by ATLAS at the LHC from $pp$ collisions at a centre-of-mass
energy of 7 TeV between March and August 2011. After application of kinematic selection criteria 
that require one top quark to decay via $W\rightarrow \ell\nu$
(where $\ell$ is either a muon or an electron)  
and identification of a jet as originating from a $b$ quark ($b$-tag), 
the dominant background to the $t \overline{t} \rightarrow \ell+ 
\tau+X$ channels with the $\tau$ lepton decaying hadronically is the $t\overline{t} \rightarrow \ell \ + $ jets channel in which a jet
is misidentified as a hadronic $\tau$ lepton decay. Therefore, $\tau$ lepton identification ($\tau$ ID) is critical for separating
signal and background.
 The $\tau$ ID methodology employed in this analysis exploits a multivariate
technique to build a discriminant ~\cite{taurec}. A boosted decision tree ~(BDT) algorithm is used \cite{BDT0}.   
The number of $\tau$ leptons in a sample is extracted by fitting the distributions of BDT outputs
to background and signal templates. The results are checked using an alternative method,
referred to as  the ``matrix method'', based on a cut on the BDT output.

\section{ATLAS Detector }
\label{s:detector}

The ATLAS detector~\cite{ATLAS} at the LHC covers nearly the entire solid angle around the collision 
point.\footnotemark[1] \footnotetext[1]{Atlas uses a right-handed coordinate system with its origin
at the nominal interaction point in the centre of the detector and the z-axis along the
beam pipe. The x-axis points to the centre of the LHC ring, and the y-axis points upwards.
The azimuthal angle $\phi$ is measured around 
the beam axis and the polar
angle $\theta$ is the angle from the beam axis. The pseudorapidity is defined as
$\eta=-\ln[\tan(\theta/2)]$. The distance $\Delta R$ in $\eta-\phi$ space is defined as 
$\Delta R=\sqrt{(\Delta \phi)^2+(\Delta \eta)^2}$.}
It consists of an inner tracking detector surrounded by a thin superconducting solenoid, 
electromagnetic (EM) and hadronic calorimeters, and an external muon spectrometer incorporating three 
large superconducting toroid magnet assemblies. The inner tracking detector provides tracking
information in a pseudorapidity range $|\eta|<2.5$. 
The liquid-argon (LAr) EM sampling calorimeters cover a range of $|\eta|<3.2$
with fine granularity. An iron--scintillator tile calorimeter provides hadronic energy
measurements in the central rapidity range ($|\eta|<1.7$). The endcap and forward regions
are instrumented with LAr calorimeters for both EM and hadronic energy measurements
covering $|\eta|<4.9$. The muon spectrometer
provides precise tracking information in a range of $|\eta|<2.7$.

ATLAS uses a three-level trigger system to select events. The level-1 trigger is implemented in hardware
using a subset of detector information to reduce the event rate below 75 kHz. This is followed by two software
based-trigger levels, level-2 and the event filter, which together reduce the event rate to about 300~Hz recorded
for analysis.

\section{Simulated Event Samples}
\label{s:mc}

Monte Carlo (MC) simulation samples are used to optimise selection
procedures, to calculate the signal acceptance and to evaluate 
contributions from some background processes. 

For the $t\overline{t}$ and single top-quark 
final states, the next-to-leading-order (NLO) generator MC@NLO~\cite{MCNLO} is used
with a top-quark mass of 172.5~GeV and with the NLO parton
distribution function (PDF) set CTEQ6.6~\cite{CTEQ}. The
``diagram removal scheme'' is used to remove overlaps between the single top-quark 
and the $t\overline{t}$ final states. The $t\overline{t}$ cross section
is normalised to the prediction of HATHOR ($164^{+11}_{-16}$~pb)~\cite{HATHOR},
 which employs an approximate
next-to-next-to-leading-order (NNLO) perturbative QCD calculation.
  
For the background channels,
MC samples of $W/Z$, single top-quark  
events and diboson $WW$, $WZ$, and $ZZ$  events (all in association with jets)
  are
used. $W$+jets events and $Z/\gamma^{*}$+jets
events (with dilepton invariant mass $m_{\ell^{+}\ell^{-}}$ $>$ 40~GeV)  are
generated by the ALPGEN generator~\cite{Alpgen} with up to five outgoing partons from the
hard scattering process, in addition to the vector bosons.\footnotemark[2]
\footnotetext[2]{The fraction of events with $m_{\ell^{+}\ell^{-}}$ $<$ 40~$\gev$ is estimated to be less than 0.2\%
of the total after all selections.} 
The MLM matching scheme of the ALPGEN
generator is used to remove overlaps between matrix-element
and parton-shower products. Parton evolution
and hadronisation is handled by HERWIG~\cite{HERWIG}, as is the generation of diboson events.
The leading-order PDF set
CTEQ6L is used for all backgrounds described above.  


All samples that use HERWIG for parton shower evolution and
hadronisation rely on JIMMY~\cite{JIMMY} for the underlying event
model.  The $\tau$-lepton decays are handled by TAUOLA \cite{TAUOLA}.
The effect of multiple $pp$ interactions per bunch crossing (``pile-up'') is modelled by overlaying simulated 
minimum bias events over the original hard-scattering event \cite{MCtune}. MC events are then reweighted so that the 
distribution of interactions per crossing in the MC simulation matches that observed in data.
The average number of pile-up events in the sample is 6.3.
After event generation, all samples
are processed with the GEANT4 ~\cite{Geant4} simulation of the ATLAS detector, the trigger simulation and 
are then subject to the same reconstruction algorithms as the data~\cite{{Atlas_sim}}.

\section{Data and Event Selection}
\label{s:object}

The event selection uses the same object definition
as in the $t \overline{t}$  cross-section measurement in the dilepton channel ~\cite{ATLAS700_dil} with 
the exception of a 
$\tau$ candidate instead of a second electron or muon candidate and some minor adjustments. The 
 electrons must be isolated and have
 $E_{\text{T}}>25\gev$ and
$|\eta_{\text{cluster}}|<2.47$, excluding the barrel-endcap transition region
($1.37<|\eta_{\text{cluster}}|<1.52$), where $E_\text{T}$ is the transverse energy and $\eta_{\text{cluster}}$ is the
pseudorapidity of the calorimeter energy cluster associated with the
candidate.
 The electron is defined as isolated if the  
$E_{\text{T}}$ deposited in the calorimeter and not associated with the
electron in a cone in $\eta$-$\phi$ space of radius $\Delta R=0.2$ is less than
$4\gev$.
The muons must also be isolated and have $\pt >20\gev$ and $|\eta|<2.5$.
For isolated muons, both the corresponding  $E_{\text{T}}$ and the analogous
track isolation transverse momentum ($\pt$) must be less than 4~GeV in a cone
of $\Delta R=0.3$. The track isolation $\pt$ is calculated from the sum
of the track transverse momenta for tracks with $\pt>1\gev$ around
the muon.
Jets are reconstructed with the anti-$k_t$ algorithm ~\cite{anti_kt} with a radius 
parameter $R=0.4$, starting from energy deposits (clusters) in the calorimeter reconstructed using
the scale established for electromagnetic objects. These jets are then calibrated to the
hadronic energy scale using $\pt$- and $\eta$-dependent correction
factors obtained from simulation ~\cite{JES2}.
The jet candidates are required to have 
$\pt>25 ~\gev$ and $|\eta|<2.5$. Jets identified as originating from a
$b$ quark ($b$-tag) by a vertex tagging algorithm are those
that pass a decay length significance cut corresponding to an efficiency of 70\%  for $b$-quark 
jets from $t \overline{t}$ 
events
and a 1\% efficiency for light-quark and gluon jets ~\cite{{ATLAS700_dil},{combNN}}.

 The missing transverse momentum 
is constructed from the vector
sum of all calorimeter cells with $|\eta|<4.5$, projected onto the
transverse plane. Its magnitude is denoted $\MET$. The hadronic energy scale is used for the energies of cells associated with jets; $\tau$ candidates are treated as jets. 
Contributions
from cells associated with electrons employ the electromagnetic energy calibration.
Contributions from the $\pt$ of muon tracks are included,
removing the contributions of any calorimeter cells associated with
the muon.

\subsection{$\tau$ Reconstruction and Identification}
\label{sec:taurec}

The reconstruction and identification of hadronically decaying $\tau$ leptons proceeds 
as follows:
\begin{enumerate} 
\item the $\tau$ candidate reconstruction starts by considering each jet 
as a $\tau$ candidate;
\item energy clusters in the calorimeter associated with the $\tau$ candidate are 
used to calculate kinematic quantities (such as $E_{\text{T}}$) and the associated tracks are found;
\item  identification variables are calculated from 
the tracking and calorimeter information; 
\item these variables are combined 
into multivariate discriminants and the outputs of the discriminants
are used to separate jets and 
electrons misidentified as $\tau$ leptons decaying hadronically from $\tau$ leptons. 
\end{enumerate}
Details are given in Ref.~9. In this analysis the outputs of
BDT discriminants are used.

Reconstructed $\tau$ candidates are required to have 
$20\gev <E_{\text{T}}<100\gev$. They must also have
$|\eta|<2.3$, and one, two or three associated
tracks. A track is associated with the $\tau$ candidate if it has $\pt>1\gev$ and is inside a cone
of $\Delta R<0.4$ around the jet axis.
 The associated track with highest $\pt$ must have $\pt>4\gev$.
The charge is given by the sum of the
charges of the associated tracks, and is required  to be non-zero.
The probability of misidentifying the $\tau$ lepton charge sign is about 1\%. 
The charge misidentification rate for muons and electrons is negligible.

  If the $\tau$ candidate overlaps with a muon ($\pt>4\gev$, no isolation
required) or an electron candidate within $\Delta R(\ell,\tau)<0.4$,
 the $\tau$ candidate is removed.  To remove electrons misidentified as
$\tau$ leptons, an additional criterion is used that relies on a BDT
trained to separate $\tau$ leptons and electrons (BDT$_e$)
using seven variables shown to be well
modelled by comparing
$Z\rightarrow e^{+}e^{-}$ and $Z\rightarrow \tau^{+} \tau^{-}$ events in data and in MC simulation.
The variables were chosen after ranking
a large set by their effectiveness.\footnotemark[3]
\footnotetext[3]{The effectiveness is quantified by quadratically summing over the change in the purity between the mother and daughter leaves for every node in which the given variable is used in a decision tree.}
The most effective variables for BDT$_e$ 
are $E/p$, the EM fraction (the ratio of the $\tau$ candidate energy measured in the EM calorimeter to the
total $\tau$ candidate energy measured in the calorimeter), and the cluster-based shower width.
The BDT output tends to be near 1 (0) if the $\tau$ candidate is a $\tau$ lepton (electron).  
The $\tau$ candidate is required to satisfy BDT$_e>0.51$; 85\% of reconstructed 
$\tau$ leptons decaying hadronically satisfy 
that requirement as measured in $Z\rightarrow\tau^{+}\tau^{-}$ events.
The additional rejection for electrons is a factor of 60.

The majority of objects reconstructed as $\tau$ candidates in a 
multi-jet environment 
are  jets misidentified as $\tau$ leptons (fake $\tau$).  
Another BDT (BDT$_j$) based on eight variables is used to separate 
$\tau$ leptons in $\tau$ candidates with one track (denoted $\tau_1$) from such jets.
For candidates with more than one track  (denoted $\tau_3$) BDT$_j$
includes ten variables. 
The most effective variables for BDT$_j$ 
 are calorimeter and track isolation, cluster-based jet 
mass, and the fraction of energy within $\Delta R=0.1$ of the jet axis.
 The BDT$_j$ distributions are fit with templates for background and signal to extract
the number of $\tau$ leptons in the sample. 
Details are given in Section ~\ref{sec:datafits}. 
The fake $\tau$ background  
in the $\tau_3$ sample is significantly higher than  
in the $\tau_1$ sample, 
leading to very different BDT$_j$ distributions. Hence independent measurements
are carried out  for $\tau_1$  and $\tau_3$ candidate events and the
results are combined at the end. If there is a  
$\tau_1$ and a $\tau_3$ candidate in the event, the $\tau_1$ candidate is kept as the
probability that the $\tau_1$ is a $\tau$ lepton is much higher. 
If there are two $\tau_1$ or $\tau_3$ candidates, both are kept.

\subsection{Event Selection}
\label{sec:selection}

For this analysis, events are selected using a single-muon trigger with a $\pt$
threshold of 18 GeV or a single-electron trigger with a $\pt$ threshold of 20 GeV, rising to 22 GeV during 
periods of high instantaneous luminosity.
The offline requirements are based on data quality criteria and optimised using Monte Carlo simulation:

\begin{itemize}
\item  a primary vertex with at least five
tracks, each with $\pt>400\mev$, associated with it;
\item one and only one isolated high-$\pt$ muon and no identified electrons for the $\mu+\tau$ channel, or 
one and only one isolated electron and no isolated muons for the $e+\tau$ channel;
\item at least one $\tau$ candidate (as defined in Section ~\ref{sec:taurec});
\item at least two jets not overlapping 
with a $\tau$ candidate, i.e. $\Delta R(\tau,\text{jet})>0.4$; 

\item $\MET>30$ GeV to reduce the multi-jet background, and
the scalar 
sum of the $\pt$ of the leptons (including $\tau$), jets, and $\MET$ must be greater then $200\gev$, to reduce 
the $W$+jets background. 
\end{itemize}

The $\ell+\tau$ samples are divided into events with
no jets identified as a $b$-quark jet (0 $b$-tag control sample) and those with 
at least one such jet ($\geq 1~b$-tag $t \overline{t}$ sample).
The 0 $b$-tag sample is used to estimate the background in the $\geq 1~b$-tag $t \overline{t}$ sample.
Each sample is split into two, one with the $\tau$ candidate and 
$\ell$ having the opposite sign charge (OS), and the other one with $\tau$
and $\ell$ having the same sign charge (SS). 
While the $\tau$ candidates in the SS samples are almost all  
fake $\tau$ leptons, the OS samples have a mixture of $\tau$ leptons and fake $\tau$ leptons.
The numbers of observed and expected events in the above samples are shown in Table~\ref{table:SelectedEvents}. 
All processes contribute more events to OS than SS because of the correlation between a leading-quark charge and
the lepton charge, 
except for the multi-jet
channel contribution which has equal number of OS and SS events within the uncertainties.
The $\ell$+jets entry includes the contribution from events with $\tau$ leptons
when the $\tau$ candidate is actually a fake $\tau$.
The $\tau$ entries require the reconstructed $\tau$ candidate be matched to a generated $\tau$ lepton. 
The matching criterion is $\Delta R<0.1$ between the $\tau$ candidate and the observable component of
the generated $\tau$ lepton.

To estimate the multi-jet background from data, an
event selection identical to the $\mu+\tau$ ($e+\tau$)
event selection except for an inverted muon (electron) isolation cut is used
to obtain a multi-jet template for the shape of the transverse mass, $m_{\text{T}}$.\footnotemark[4]
\footnotetext[4]{$m_{\text{T}}=\sqrt{(E_T^{\ell}+\MET)^2-(p_x^{\ell}+E_x^{\text{miss}})^2
-(p_y^{\ell}+E_y^{\text{miss}})^2}$.} 
The normalization of each selected data sample is obtained by fitting
the $m_{\text{T}}$ distribution of the selected data samples
with the multi-jet template and the sum of non-multi-jet processes predicted by MC, 
allowing the amount of both to float.
The uncertainty on the multi-jet background is estimated to be 30\%. However, because of the subtraction method discussed in Section \ref{sec:background}, the multi-jet background 
plays no role in the cross-section measurement. There are small differences
 between the total number
of events predicted  and observed which motivate using data as much as possible to estimate
the background.

As one can see from Table \ref{table:SelectedEvents},
 the $\tau$ leptons are almost all in the OS sample and come mainly from 
two sources:
$Z\rightarrow \tau^{+} \tau^{-}$, which is the dominant source in the sample with 0 $b$-tag, and $t \overline{t}\rightarrow
\ell+\tau+X$ which is the dominant source in the sample $\geq 1$~$b$-tag. The sources of fake $\tau$ leptons are also
quite distinct between the 0 $b$-tag and the $\geq 1$ $b$-tag samples: 
the first is mainly $W/Z$+jets with
small contributions from other channels, the second is mainly $t \overline{t}$.

\begin{table*}[h!bt]
\begin{center}
\caption{\label{table:SelectedEvents} Number of $\ell + \tau$  candidates for Monte Carlo samples and data. 
$t \overline{t}(\ell+e)$
  are $t \overline{t}$ events with one identified lepton and an  electron reconstructed as a $\tau$ candidate. 
$t \overline{t}(\ell+\text{jets})$
  are $t \overline{t}$  events with one identified lepton and a jet reconstructed as a $\tau$ candidate.
$\ell$+jets are events with one identified lepton and  a jet reconstructed as a $\tau$ candidates from sources other
than $t \overline{t}$ and multi-jets. Sources contributing to jet fakes are $W$+jets, $Z$+jets, single top-quark and diboson events.
$Wt(\ell+\tau)$ is $W+t$ production with one $W$ decaying to $\ell$ and another to $\tau$.
Excepting multi-jets the uncertainties are statistical only. MC samples are normalized to the data integrated luminosity}
\begin{small}
\begin{tabular}{r|rr|rr|rr|rr}
 \hline
 \hline 
$\mu+\tau$  & \multicolumn{4}{|c|}{$\tau_1$} & \multicolumn{4}{|c}{$\tau_3$} \\
\hline
  & \multicolumn{2}{|c|}{0 $b$-tag} & \multicolumn{2}{|c}{$\geq 1$ $b$-tag}
& \multicolumn{2}{|c|}{0 $b$-tag} & \multicolumn{2}{|c}{$\geq 1$ $b$-tag}  \\ 
 & OS & SS  & OS & SS  & OS & SS  & OS & SS  \\ \hline
$t\overline{t} (\mu+\tau)$ & 60 $\pm$  \phantom{00}2 & $<1$ & 390 $\pm$  \phantom{0}4 & 2 $\pm$  \phantom{0}1 & 17 $\pm$  \phantom{00}1 & 1 $\pm$  \phantom{00}1 & 118 $\pm$\phantom{00}3  & 2 $\pm$  \phantom{00}1 \\ \hline 
$t\overline{t} (\mu+e$) & 3 $\pm$  \phantom{00}1 & $<1$ & \phantom{0}12 $\pm$  \phantom{0}1 & 1 $\pm$  \phantom{0}1 & 1 $\pm$  \phantom{00}1 & $<1$ & 3 $\pm$  \phantom{00}1 & $<1$   \\ 
$t\overline{t} (\mu+$jets) & 308 $\pm$  \phantom{00}4 & 163 $\pm$  \phantom{00}3 & 1528 $\pm$  \phantom{0}9 & 660 $\pm$  \phantom{0}6 & 685 $\pm$  \phantom{00}6 & 443 $\pm$  \phantom{00}5 & 3484 $\pm$  \phantom{0}13 & 2000 $\pm$  \phantom{0}10   \\ 
$\mu+$jets & 5010 $\pm$ \phantom{0}70 & 3020 $\pm$  \phantom{0}60 &  \phantom{0}496 $\pm$ 17 & 297 $\pm$ 13 & 12230 $\pm$ 120 & 8670 $\pm$  \phantom{0}90 & 1293 $\pm$  \phantom{0}28 & 928 $\pm$  \phantom{0}24   \\ 
Multi$-$jets &  \phantom{0}470 $\pm$ 140 &  \phantom{0}540 $\pm$ 160 &  \phantom{0}117 $\pm$ 35 & 150 $\pm$ 40 & 990 $\pm$ 300 & 1120 $\pm$ 340 & 460 $\pm$ 140 & 400 $\pm$ 120  \\ 
$Wt (\mu+\tau)$ & 7 $\pm$  \phantom{00}1 & $<1$ & \phantom{0}18 $\pm$  \phantom{0}1 & \phantom{0}1 $\pm$  \phantom{0}1 & 2 $\pm$  \phantom{00}1 & $<1$ & 5 $\pm$  \phantom{00}1 & $<1$  \\ 
$Z\rightarrow \tau\tau$ & 301 $\pm$  \phantom{0}13 & 2 $\pm$  \phantom{00}1 & \phantom{0}16 $\pm$  \phantom{0}3 & $<1$ & 75 $\pm$  \phantom{00}7 & 1 $\pm$  \phantom{00}1 & 3 $\pm$  \phantom{00}2 & $<1$  \\ 
\hline
Total & 6160 $\pm$ 160 & 3730 $\pm$ 170 & 2580 $\pm$ 40 & 1110 $\pm$ 40 & 14000 $\pm$ 320 & 10230 $\pm$ 350 & 5370 $\pm$ 140 & 3330 $\pm$ 120  \\ 
Data & \multicolumn{1}{c}{5450}  & \multicolumn{1}{c|}{3700}  & \multicolumn{1}{c}{2472} & \multicolumn{1}{c|}{1332}  & \multicolumn{1}{c}{13322}  & \multicolumn{1}{c|}{10193}  & \multicolumn{1}{c}{5703}  & \multicolumn{1}{c}{3683}  \\ 
\hline
 \hline 
$e+\tau$  & \multicolumn{4}{|c|}{$\tau_1$} & \multicolumn{4}{|c}{$\tau_3$} \\
\hline
  & \multicolumn{2}{|c|}{0 $b$-tag} & \multicolumn{2}{|c}{$\geq 1$ $b$-tag}
& \multicolumn{2}{|c|}{0 $b$-tag} & \multicolumn{2}{|c}{$\geq 1$ $b$-tag}  \\ 
 & OS & SS  & OS & SS  & OS & SS  & OS & SS  \\ \hline
$t\overline{t} (e+\tau)$ & 54 $\pm$  \phantom{00}7 & 1 $\pm$  \phantom{00}1 & 342 $\pm$ 19 & 3 $\pm$  \phantom{0}2 & 15 $\pm$  \phantom{00}4 & $<1$ & 103 $\pm$  \phantom{0}10 & 2 $\pm$  \phantom{00}1   \\ \hline
$t\overline{t} (e+e$) & 2 $\pm$  \phantom{00}1 & $<1$ & 11 $\pm$  \phantom{0}3 & 1$\pm$ \phantom{00}1 & $<1$ & $<1$ & 2 $\pm$  \phantom{00}1 & $<1$  \\ 
$t\overline{t} (e+$jets) & 273 $\pm$  \phantom{0}17 & 146 $\pm$  \phantom{0}12 & 1340 $\pm$ 40 & 599 $\pm$ 25 & 633 $\pm$  \phantom{0}25 & 399 $\pm$  \phantom{0}20 & 3090 $\pm$  \phantom{0}60 & 1780 $\pm$  \phantom{0}40   \\
$e+$jets & 3950 $\pm$  \phantom{0}60 & 2590 $\pm$  \phantom{0}50 & 380 $\pm$ 20 & 256 $\pm$ 16 & 10140 $\pm$ 100 & 7530 $\pm$  \phantom{0}90 & 1120 $\pm$  \phantom{0}33 & 841 $\pm$  \phantom{0}29   \\ 
Multi$-$jets & 600 $\pm$ 180 & 620 $\pm$ 190 & 170 $\pm$ 50 & 140 $\pm$ 40 & 2000 $\pm$ 600 & 2000 $\pm$600 & 690 $\pm$ 210 & 610 $\pm$ 180  \\ 
Z $\rightarrow ee$ & 92 $\pm$  \phantom{0}10 & 3 $\pm$  \phantom{00}2 & 9 $\pm$  \phantom{0}3 & $<1$ & 11 $\pm$  \phantom{00}3 & 2 $\pm$  \phantom{00}1 & $<1$ & $<1$  \\ 
$Wt (e+\tau)$ & 7 $\pm$  \phantom{00}3 & $<1$ & 17 $\pm$  \phantom{0}4 & $<1$ & 1 $\pm$  \phantom{00}1 & $<1$ & 5 $\pm$  \phantom{00}2 & $<1$   \\ 
$Z\rightarrow \tau\tau$ & 217 $\pm$  \phantom{0}15 & 2 $\pm$  \phantom{00}1 & 15 $\pm$  \phantom{0}4 & $<1$ & 60 $\pm$  \phantom{00}7 & 1 $\pm$  \phantom{00}1 & 3 $\pm$  \phantom{00}2 & $<1$   \\ 
\hline
Total & 5190 $\pm$ 190 & 3360 $\pm$ 200 & 2280 $\pm$ 70 & 990 $\pm$ 50 & 12900 $\pm$ 600 & 9900 $\pm$ 600 & 5020 $\pm$ 220 & 3230 $\pm$ 180  \\ 
Data & \multicolumn{1}{c}{5111}  & \multicolumn{1}{c|}{3462} & \multicolumn{1}{c}{2277}  & \multicolumn{1}{c|}{1107}  & \multicolumn{1}{c}{12102}  & \multicolumn{1}{c|}{9635}  & \multicolumn{1}{c}{5033}  & \multicolumn{1}{c}{3192}  \\ 

\hline
\hline
\end{tabular} 
\end{small}

\end{center}
\end{table*}

\section{Background Models}
\label{sec:background}

The jet origin can strongly influence the $\tau$-lepton fake probability. 
Due to their narrow shower width and lower track 
multiplicity, light-quark jets have a higher probability of faking a
$\tau$ lepton than other jet types. Thus the BDT$_j$ distributions have a strong
dependence on the jet type. It is therefore crucial to
build a background model which properly reflects the jet composition
in order to correctly estimate the fake $\tau$ contamination in the signal
region. Deriving this background model from control
regions in data rather than MC simulation is preferable in order to avoid systematic
effects related to jet composition in the MC models.

The gluon component of the fake $\tau$ leptons is charge symmetric; therefore it
is expected to have the same shape in SS events as in OS events and should
contribute the same number of fake $\tau$ leptons in each sample.
The contribution of fake $\tau$ leptons from gluons can  be removed 
by  subtracting the distribution of any quantity for SS events from the corresponding distribution 
for OS events. The multi-jet background
also cancels, as can be seen in Table \ref{table:SelectedEvents}.
The resulting distributions
are labeled OS-SS.
Similarly, since each sample is expected to have 
an almost equal contribution from $b$-jets and $\overline{b}$-jets, 
the small $b$-jet component should also 
be removed by OS-SS (asymmetric single $b$ production is negligible compared to $b \overline{b}$ production). 
The only jet types remaining in the OS-SS distributions are light-quark jets. 
MC studies indicate that the BDT$_j$ distributions of $c$-quark jets misidentified as $\tau$ leptons 
are not noticeably different from those of light-quark jets. 

One can construct a background BDT$_j$ distribution from the 0 $b$-tag 
data sample by subtracting the 
expected amount of true $\tau$ signal. The signal is  mainly from $Z\rightarrow \tau^{+} \tau^{-}$ 
and can be reliably predicted from MC.
A control sample dominated by  $W$+jets events is considered as a check.
The latter sample  is selected by requiring events with
a muon and a $\tau$ candidate, no additional jets, $\MET>30~\gev$
 and $40~\gev<m_{\text{T}}<100~\gev$.  According to MC simulation, 
in $W$+jets  events where exactly one jet is required,
 90\% of the fake $\tau$ leptons are from light-quark jets and 10\% from gluons.
This sample is labeled $W+1$ jet.

The BDT$_j$ background shapes for the OS-SS 0 $b$-tag and $\ge 1$ $b$-tag data samples are not 
identical to 
the $W+1$ jet distributions
for two reasons: (1) the shape depends on the 
jet multiplicity, (2) different OS/SS ratios are observed in the samples. 
The dependence on the OS/SS ratio comes from the differences in jet fragmentation for
a leading particle with the opposite charge and the same charge as the initial quark.
MC studies of the ratio of  
OS-SS BDT$_j$ background 
distributions derived from $W+1$ jet and $\geq 1$ $b$-tag show that 
significant corrections are needed (30\% for BDT$_j>0.8$, a region dominated by the true $\tau$ signal).
For the 0 $b$-tag sample the corresponding corrections are much smaller
(5\% in the same region).
 Both the 0 $b$-tag and the $W+1$ jet data samples are used to obtain 
statistically independent estimates of the background in the $\geq 1$ $b$-tag
sample.

Two different approaches are used for constructing backgrounds 
in the $\geq 1$ $b$-tag data sample.
One, used by the fit method (Section~\ref{sec:datafits}), is to reweight the BDT$_j$ distribution 
of the background bin-by-bin 
using the MC-based ratio of the $\geq 1$ $b$-tag background 
to the background model. In this case the 0 $b$-tag sample is preferred as it 
requires smaller corrections derived from MC simulation; 
the $W+1$ jet is used as a cross check.
The other approach is to split the 
background into bins 
of some variable within which the shapes of BDT$_j$ 
distributions  of the background model are close to those
from the $\ge 1$ $b$-tag background. This approach, 
used in the Matrix Method cross check (Section~\ref{sec:matrix}),
 avoids using MC corrections,
but assumes the data and MC simulation behave similarly as function of the binning variable.

\section{Fits to BDT$_j$ Distributions}

\label{sec:datafits}

The contribution from $t \overline{t}\rightarrow \ell+\tau+X$ 
signal is derived from the
$\geq 1$ $b$-tag data sample by  a $\chi^2$ fit to the OS$-$SS BDT$_j$ distribution
with a background template and a signal template. The parameters of the 
fit are the amount of background and the amount of signal. The shapes of the templates are fixed.

Two background templates corrected by MC, as discussed in Section ~\ref{sec:background}, are used: 
one derived from 0 $b$-tag data, 
the other from the $W+1$ jet
data sample. 
The signal BDT$_j$ templates for 0 $b$-tag and $\ge 1$ $b$-tag are derived from $\tau$ leptons in 
$t\overline{t}$ and $Z \rightarrow \tau^{+} \tau^{-}$  MC simulation. 
Contributions to the BDT$_j$ distributions from electrons passing the BDT$_e$ cut cannot be distinguished 
from $\tau$ leptons so they 
are treated as part of the signal.

The uncertainty on the background templates is determined by the
numbers of data and MC simulated events. The signal template for the 0 $b$-tag control
sample also has non-negligible statistical uncertainty (2\% for $\tau_1$, 5\% for $\tau_3$)
because of the low acceptance.

The fitting procedure was tested
extensively with MC simulation before applying it to data. 
 In the fits to the $\geq 1$ $b$-tag data, 
applying MC corrections to
the 0 $b$-tag background template increases the statistical uncertainty but raises the measured cross 
section by only 1\%.

\begin{figure}[!hbt]
\begin{center}
\begin{tabular}{cc}
\epsfig{figure=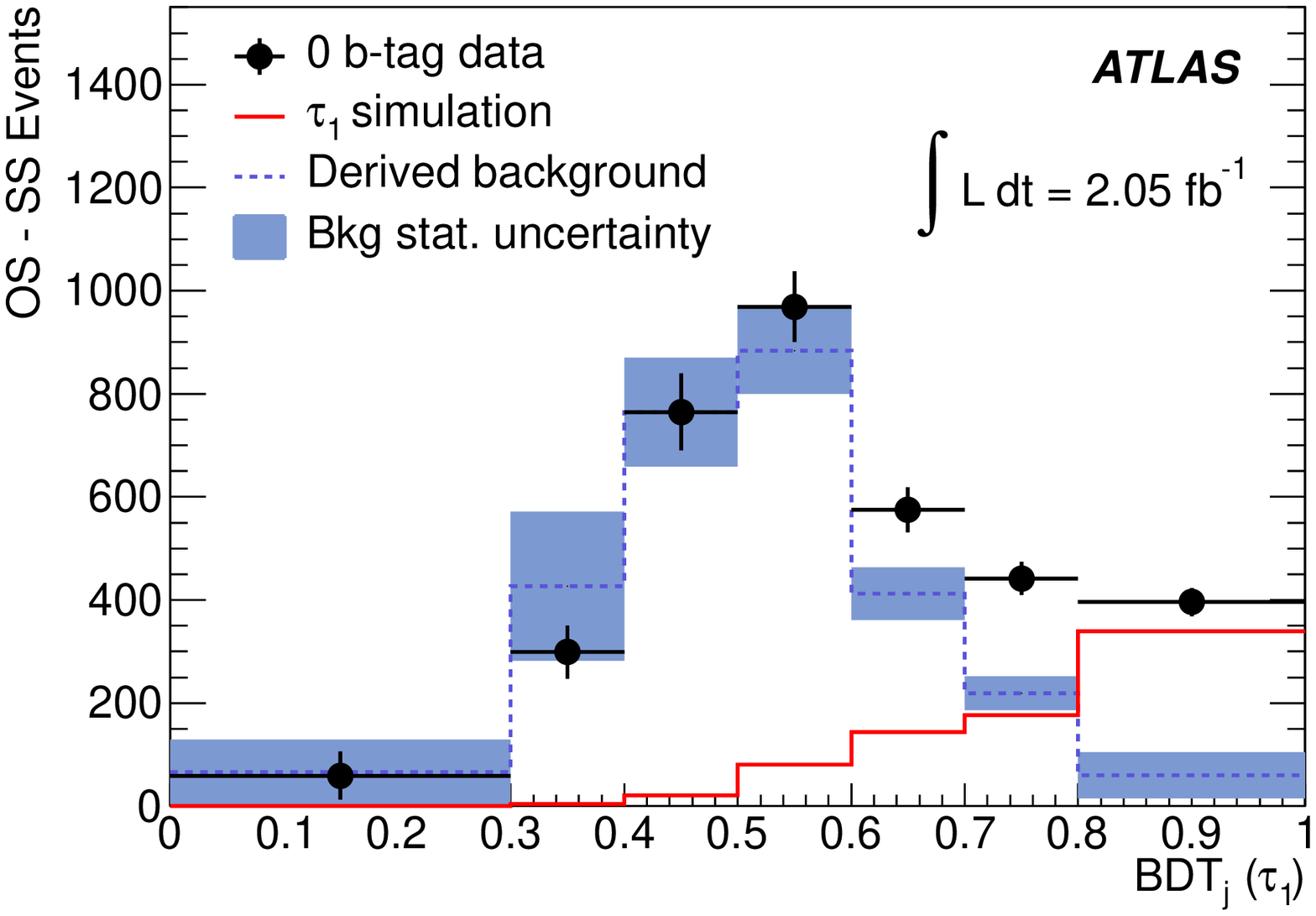,width=7.5cm} \\
\epsfig{figure=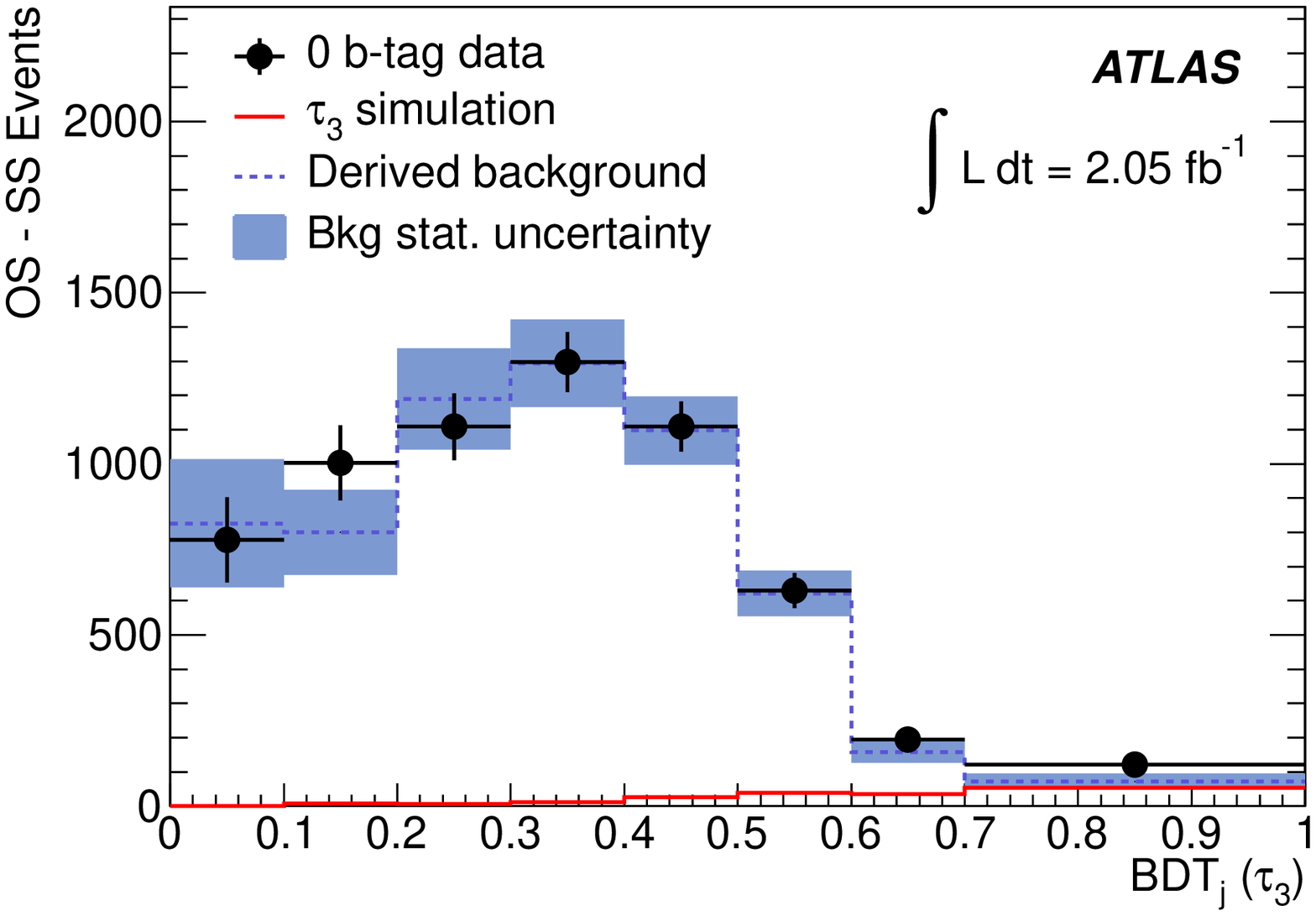,width=7.5cm} \\
\end{tabular}
\caption{\label{fig:nobtag} BDT$_j$ (OS-SS) distributions of $\ell+\tau$ ($e$ and $\mu$ combined) events in 
the 0 $b$-tag data (black points). 
The expected contributions from $\tau$ and
$e$ are shown as a solid red line.  The derived background templates 
are shown as dashed histogram with shaded/blue statistical uncertainty bands. 
The shapes of these background templates are used for the fits to the $\geq 1$ $b$-tag distributions after applying MC corrections. Top is for $\tau_1$, bottom for $\tau_3$.}
\end{center}
\end{figure}

Figure~\ref{fig:nobtag} shows the BDT$_j$ (OS-SS) distributions of $\ell+\tau$
events with 0 $b$-tag
and the 0 $b$-tag background template after subtracting the expected number of $\tau$ leptons and 
applying the MC corrections.
The $\tau$ signal is mostly $Z\rightarrow \tau^+\tau^-$ events with a
small contamination of electrons faking $\tau$ leptons (from $t \overline{t}\rightarrow \ell+e+X$ 
and $Z\rightarrow e^+e^-$) and a 
small contribution from $t\overline{t}\rightarrow \ell+\tau+X$. 
The uncertainty on the background template includes the statistical 
uncertainty of 
the correction, the statistical uncertainty from MC 
and the 0 $b$-tag data uncertainty.

\begin{figure}[!hbt]
\begin{center}
\epsfig{figure=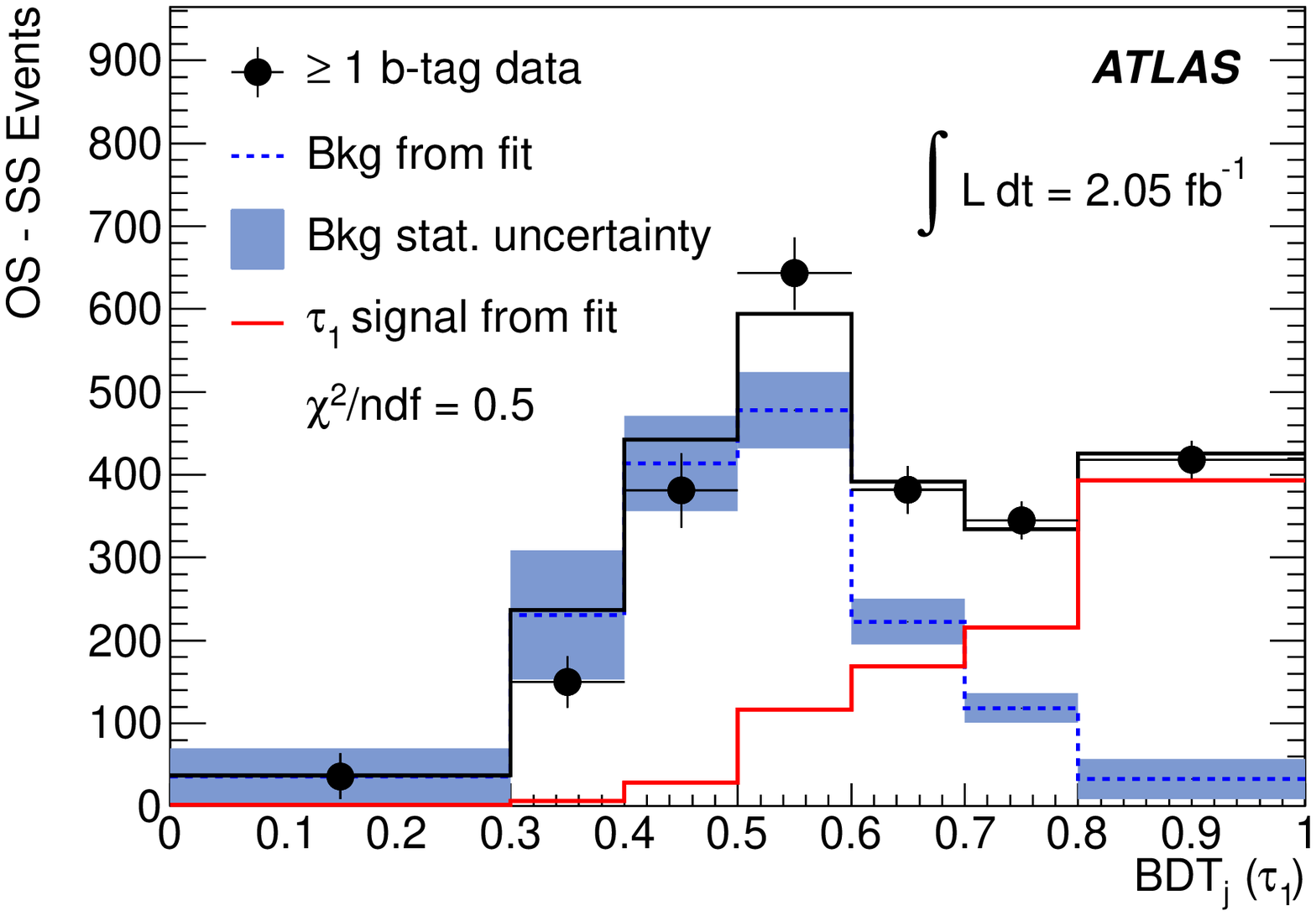,width=8.0cm} 
\\ 
\epsfig{figure=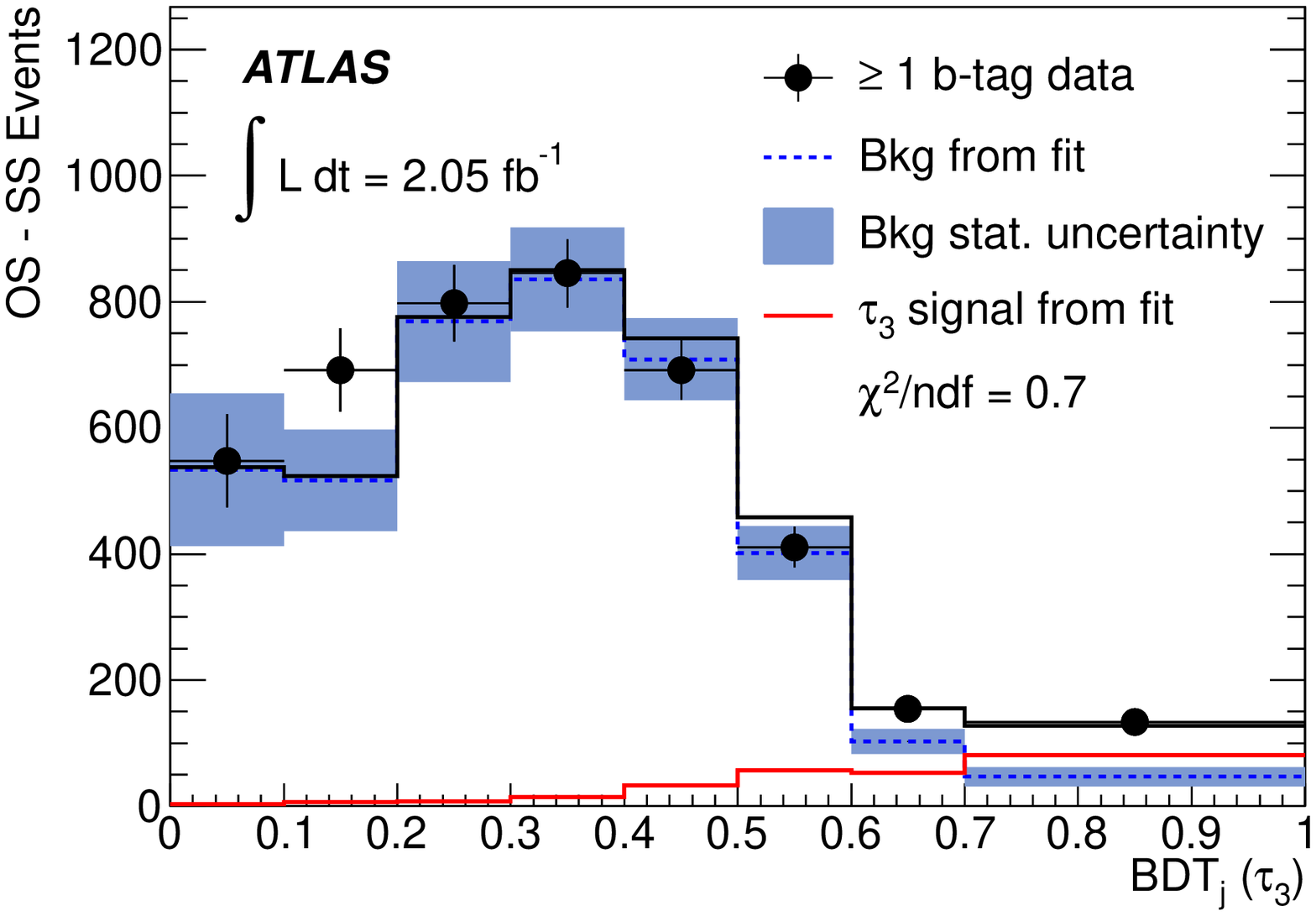,width=8.0cm}\\ 
\caption{\label{fig:fits}  BDT$_j$ (OS-SS) distributions of $\ell+\tau$ in the $\geq 1$ $b$-tag sample.
The normalisation of each template is 
derived from a fit to the data. The fitted contributions are 
 shown as the light/red (signal), dashed/blue (background derived from 0 $b$-tag after
applying MC corrections) and dark/black (total) lines. 
 Shaded/blue bands are the statistical uncertainty of the background template.}
\end{center}
\end{figure}

Figure~\ref{fig:fits} shows the result of the fit to the $\geq 1$ $b$-tag samples.
The $\tau$ lepton signal is mostly $t \overline{t}\rightarrow \ell+\tau+X$
 with a small contamination of misidentified electrons 
(estimated by applying fake probabilities derived from data),
and small contributions from $Z\rightarrow \tau^{+}\tau^{-}$ events and single top-quark events 
(estimated from MC simulation).
These contributions are subtracted from the number of signal 
events before calculating the cross section.
The fit results using the background templates derived from 0 $b$-tag data and $W+1$ jet data
are shown in Table~\ref{table:FitMethodResults}. The results are consistent with each other within
the statistical uncertainties of the background templates. The BDT$_j$ distributions for $\tau_1$ 
and $\tau_3$ are fitted separately.
The combined $\ell+\tau_i$
results are obtained by fitting the sum of the distributions.
After adding $\ell+\tau_1$ and $\ell+\tau_3$ signals obtained from 
a $\chi^2$ fit to the combined $e+\tau$ and $\mu+\tau$
distributions and subtracting the small contributions to the signal
from $Z\rightarrow\tau^{+}\tau^{-}$, $Z\rightarrow e^{+}e^{-}$ and $\ttbar\rightarrow e+\ell$
 (given in Table~\ref{table:SelectedEvents}) the results are
$840\pm70$ ($243\pm60$) $\ttbar\rightarrow\ell+\tau_1(\tau_3)+X$ events. 
The uncertainty is from the 
fit only and does not include  systematic uncertainties.
The results are in good agreement with the $780\pm50$ ($243\pm60$) events obtained with the 
$W+1$ jet background template and consistent with the number expected from
MC simulation, $726\pm19$ ($217\pm10$). Note that the fit uncertainty is 
dominated by the uncertainty on the 
background template, thus the statistical uncertainties of the results with the two different
background templates are not strongly correlated.

\begin{table}[h!bt]
\caption{\label{table:FitMethodResults} Results of template fits to $\mu+\tau$, $e+\tau$ and the combined
BDT$_j$ distributions. The combined results
are obtained by fitting the sum of the $\mu+\tau$ and $e+\tau$ BDT$_j$ distributions.
The first column gives the channel and the second the $\tau$ type.
The third column shows the extracted signal (sum of $\tau$ leptons and electrons misidentified as $\tau$ leptons) with the background
template derived from 0 $b$-tag data distributions. The fourth column shows the extracted signal
with the background template derived from $W+1$ jet. The uncertainties are from the uncertainties in
the fit parameters and do not include the systematic uncertainties. 
The MC columns give the expected $\tau$ signal and the expected number of $t \overline{t}\rightarrow l+\tau$ 
events after subtracting the contribution from non-$t \overline{t}$ events to the signal,
assuming the theoretical $t \overline{t}$ cross section (164 pb).}
\begin{center}
\begin{tabular}{l|c|cc|cc} 
\hline
\hline
& & \multicolumn{2}{c|}{Background template} & \multicolumn{2}{|c} {MC} \\ 
&  & 0 $b$-tag  & $W+1$ jet  & Signal & $t \overline{t}$ \\
\hline 
$\mu+\tau$ & $\tau_1$ & 490 $\pm$ 40  & 456 $\pm$ 32  & 432 & 388 \\
 & $\tau_3$& 135 $\pm$ 33   & 130 $\pm$ 50   & 126 & 116\\
\hline
$e+\tau$ & $\tau_1$ & 440 $\pm$ 50   & 430 $\pm$ 50   & 388 & 338 \\
 & $\tau_3$& 116 $\pm$ 32   & 120 $\pm$ 28   & 114 & 101\\
\hline
Combined & $\tau_1$ & 930 $\pm$ 70   & 860 $\pm$ 50   & 820 & 726\\
 & $\tau_3$& 260 $\pm$ 60   & 260 $\pm$ 40  & 239 & 217\\
\hline\hline
\end{tabular}

\end{center}
\end{table}

\begin{figure}[!hbt]
\begin{center}
\epsfig{figure=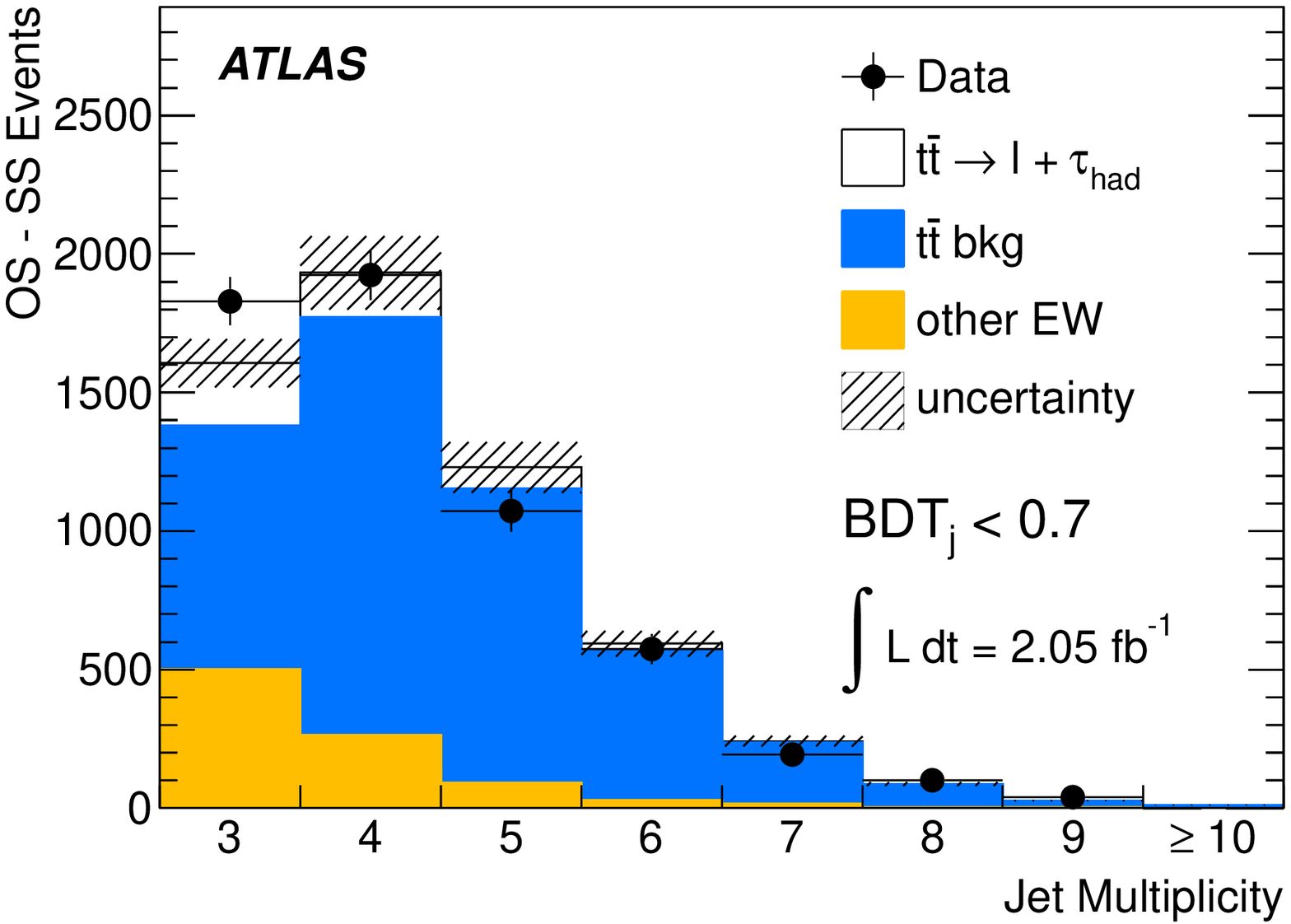,width=8.0cm} \\
\epsfig{figure=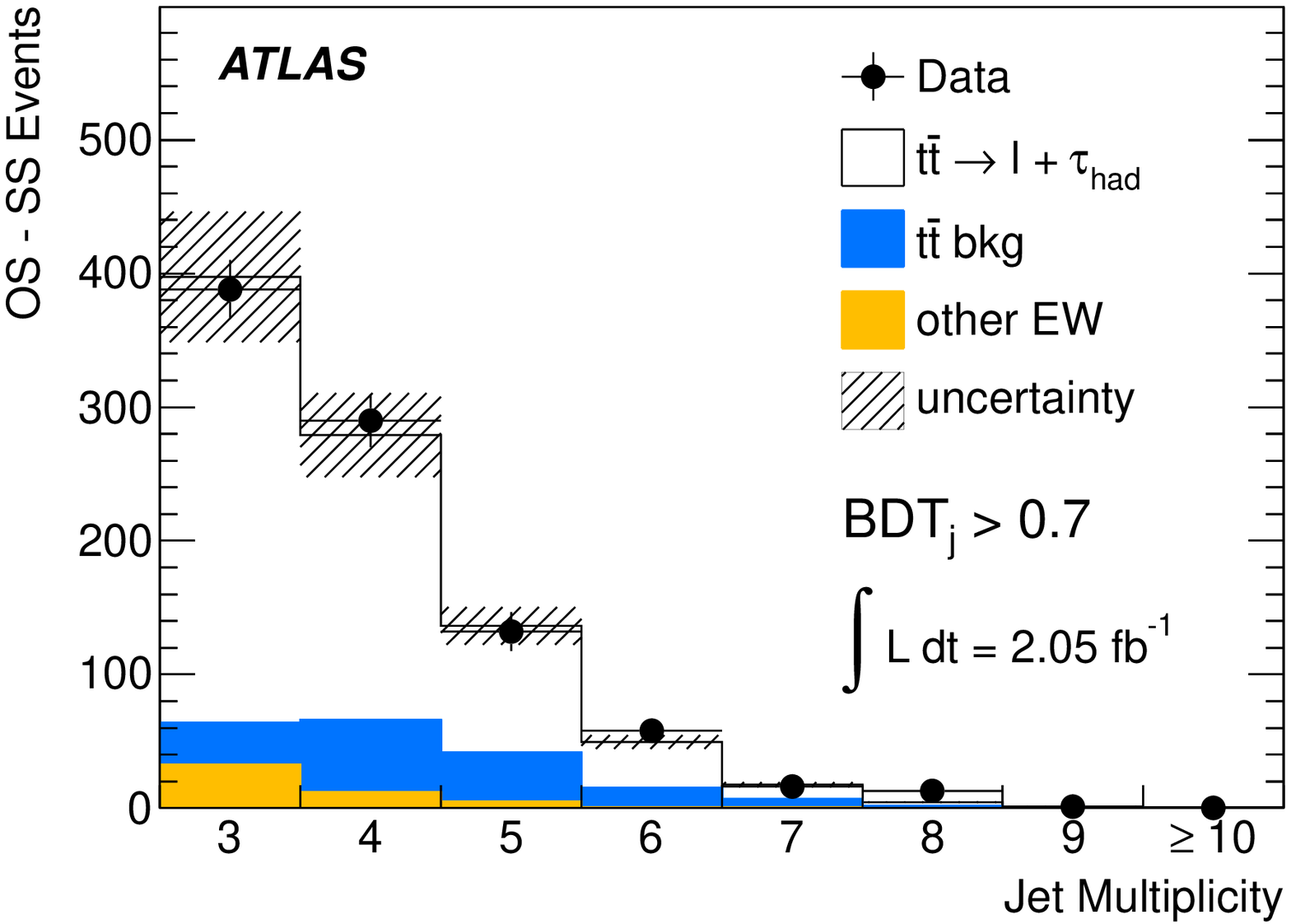,width=8.0cm} \\
\caption{\label{fig:Njets}  OS-SS number of jets distributions for events with at least one $b$-tag.
The $\mu+\tau$ and $e+\tau$ channels have been summed together. 
The solid circles indicate data and the histograms 
indicate the expected signal and backgrounds. 
The normalisation of the expected signal and the backgrounds are based on the fit result. 
The uncertainty includes statistical and systematic contributions. 
The fraction of each background is estimated from MC.
Top is for BDT$_j<0.7$, bottom for BDT$_j>0.7$.}
\end{center}
\end{figure}

\begin{figure}[!hbt]
\begin{center}
\epsfig{figure=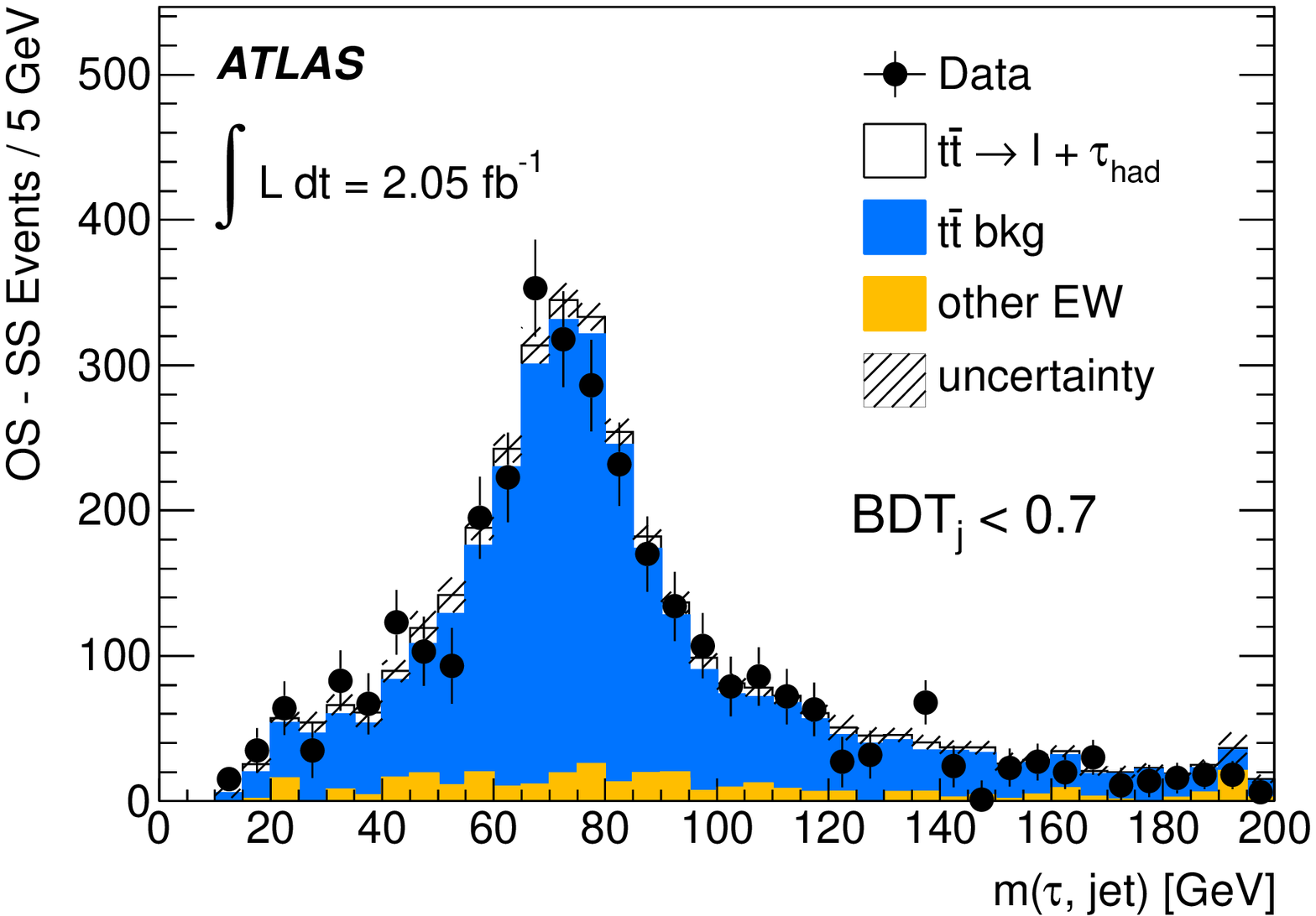,width=8.0cm} \\
\epsfig{figure=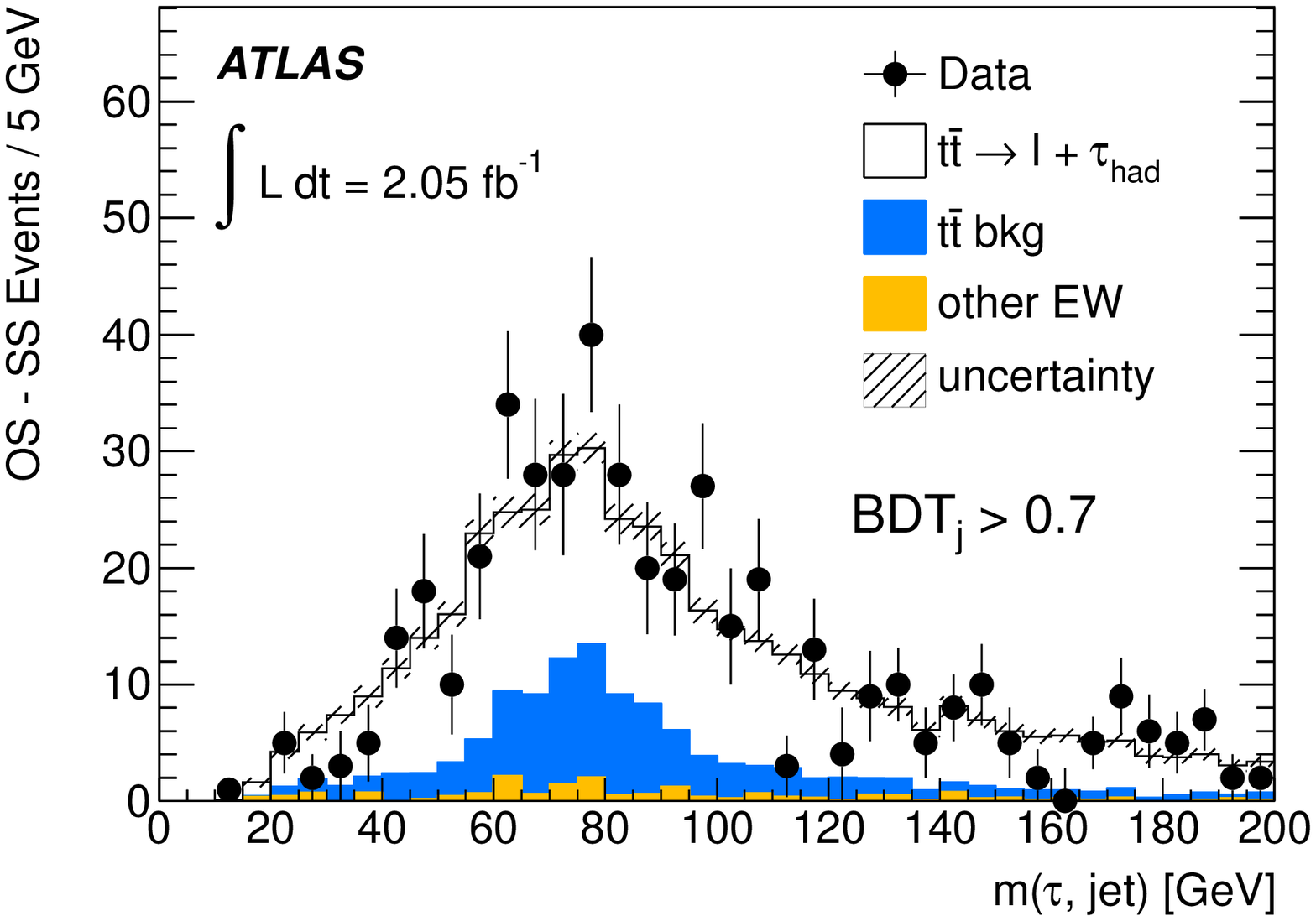,width=8.0cm} \\
\caption{\label{fig:mjtau}  OS-SS invariant mass of jet and $\tau$ candidate for events with at least one $b$-tag.
The jet is the highest $p_T$ untagged jet in events with more than one $b$-tag and the second highest $p_T$
untagged jet in events with one $b$-tag. 
The $\mu+\tau$ and $e+\tau$ channels have been summed together. 
The solid circles indicate data and the histograms 
indicate the expected signal and backgrounds. 
The normalisation of the expected signal and the backgrounds are based on the fit result. 
The uncertainty includes statistical and systematic contributions.  
The fraction of each background is estimated from MC.
Top is for BDT$_j<0.7$, bottom for BDT$_j>0.7$.}
\end{center}
\end{figure}

Figure ~\ref{fig:Njets} shows the OS-SS distribution of the number of jets for $\ge 1$ $b$-tag events 
adding all channels for two BDT$_j$ regions:
BDT$_j<0.7$, which is dominated by $t \overline{t} \rightarrow \ell \ + $ jets,
and BDT$_j>0.7$, in which the largest contribution is from
$t \overline{t} \rightarrow \ell+\tau+X$. As expected, the multiplicity of
jets peaks at four when BDT$_j<0.7$ and three when BDT$_j>0.7$ (the $\tau$
is counted as a jet). 
Figure \ref{fig:mjtau} shows the invariant mass of a selected jet with the $\tau$ candidate
for BDT$_j<0.7$ and BDT$_j>0.7$ for events with a $\tau$ candidate and three or more jets.
The selected jet is the highest $\pt$ untagged jet in events with more than one $b$-tag and the second highest $\pt$
untagged jet in events with one $b$-tag.
 The distribution shows
clearly the presence of a $W$ decaying to two jets in the BDT$_j<0.7$ region 
dominated by $t \overline{t} \rightarrow \ell \ + $ jets. The mass distribution 
in the BDT$_j>0.7$ signal region is significantly broader as expected for
$t \overline{t}\rightarrow \ell+\tau+X$. The signal and background shown in these figures are based
on the fit using the 0 $b$-tag background template.

\subsection{Check with Matrix Method}

\label{sec:matrix}
From Figures \ref{fig:Njets} and \ref{fig:mjtau} one can see that a BDT$_j>0.7$ requirement separates
well a region dominated by $t \overline{t}\rightarrow \ell$+jets from a region dominated by
$t \overline{t}\rightarrow \ell+\tau+X$.
One can extract the signal from the same OS-SS $\geq 1$ $b$-tag sample
used by the fit method via a matrix method. 
All $\tau$ candidates are labeled ``loose'', and $\tau$ candidates with BDT$_j>0.7$ are 
labeled ``tight''. The probability that the loose $\tau$
candidates are also tight $\tau$ candidates, for both $\tau$ leptons and fake $\tau$ leptons, is defined as 
\begin{equation*}
\epsilon_{\text{real}}=\frac{N^{\text{tight}}_{\text{real}}}{N^{\text{loose}}_{\text{real}}}~~~~~\epsilon_{\text{fake}}=\frac{N^{\text{tight}}_{\text{fake}}}{N^{\text{loose}}_{\text{fake}}}
\end{equation*}
where the ``real'' subscript denotes  $\tau$ lepton,  the ``fake'' subscript denotes 
fake $\tau$ and $N$ is the number of $\tau$ candidates. 
The number of ``tight'' $\tau$ leptons is then given by
\begin{equation*}
N^{\text{tight}}_{\text{real}}=N^{\text{tight}}_{\text{data}}-\frac{\epsilon_{\text{fake}}}{\epsilon_{\text{real}}-\epsilon_{\text{fake}}}(N^{\text{loose}}_{\text{data}}\cdot \epsilon_{\text{real}}-N^{\text{tight}}_{\text{data}}).
\end{equation*}
The value of $\epsilon_{\text{fake}}$ is measured utilizing the 
OS-SS BDT$_j$ distributions from the background control samples; 
$\epsilon_{\text{real}}$ is derived from MC and was tested using $Z\rightarrow \tau^{+} \tau^{-}$ events.
This method uses the binning approach described in Section \ref{sec:background} to estimate the background.
Values of $\epsilon_{\text{fake}}$ and $\epsilon_{\text{real}}$ are measured separately for three EM-fraction bins.
The EM-fraction, 
the ratio of the energy measured in the EM calorimeter to the total $\tau$ candidate energy 
measured in the calorimeter, is an 
effective variable for splitting the
data into regions where the shapes of MC OS-SS BDT$_j$ distributions for the $W$+1 jet background
template and the $\ge 1$ $b$-tag background are similar.
Table \ref{MM_btag} shows the number of signal events obtained with the matrix method using the background
derived from the 0 $b$-tag data sample and from the $W+1$ jet data sample. The numbers in each pair are in good 
agreement and consistent with the numbers obtained by fitting the OS-SS BDT$_j$ distributions.

\begin{table}[t]
\caption{\label{MM_btag} Number of signal events obtained with the matrix method for 
$\mu+\tau$, $e+\tau$ and the combined 
channels. The first column gives the channel and the second the $\tau$ type.
The third column shows the extracted signal with the background
template derived from 0 $b$-tag data distributions. The fourth column shows the extracted signal
with the background template derived from $W+1$ jet.
 In order to
compare the matrix method results to the fit results the number of signal
events shown is $\sum N^{\text{real}}_{\text{tight}}/\bar \epsilon_{\text{real}}$ where $\bar \epsilon_{\text{real}}$
is the $\epsilon_{\text{real}}$ averaged over the three EM-fraction bins. The uncertainties are statistical only.}

\begin{center}
\begin{tabular}{l|c|cc} 
\hline
\hline
& & \multicolumn{2}{c}{Background template} \\
&  & 0 $b$-tag  & $W+1$ jet \\
\hline 
$\mu+\tau$ & $\tau_1$ & 460 $\pm$ 50 & 440 $\pm$ 50 \\
 & $\tau_3$& 130 $\pm$ 40 & 105 $\pm$ 35 \\
\hline
$e+\tau$ & $\tau_1$ & 420 $\pm$ 60  & 350 $\pm$ 50 \\
 & $\tau_3$& 140 $\pm$ 40 & 160 $\pm$ 40 \\
\hline
Combined & $\tau_1$ &  880 $\pm$ 70 & 800 $\pm$ 70 \\
 & $\tau_3$& 270 $\pm$ 60 & 260 $\pm$ 60 \\
\hline\hline
\end{tabular}

\end{center}
\end{table}

\subsection{Systematic Uncertainty}
\label{sec:fits-systematics}
Several experimental and theoretical sources of systematic  uncertainty are considered.
Lepton trigger, reconstruction and selection efficiencies are assessed by comparing the
$Z \rightarrow\ell^{+}\ell^{-}$ events selected with the same object criteria
as used for the $t \overline{t}$ analyses in data and MC.

Scale factors are applied to MC samples when calculating
acceptances to account for any differences between predicted and observed efficiencies.
The scale factors are evaluated by comparing the observed efficiencies with those determined 
with simulated $Z$ boson events. 
Systematic uncertainties on these scale factors are evaluated by varying 
the selection of events 
used in 
the efficiency measurements and by checking the stability of the measurements 
over the course of data taking.
	
The modeling of the lepton momentum scale and resolution is studied using
reconstructed invariant mass distributions of $Z \rightarrow \ell^{+}\ell^{-}$\ candidate 
events and used to adjust the simulation accordingly \cite{atlase,atlasmu}.

The jet energy scale (JES) and its uncertainty are derived by
combining information from test-beam data, LHC collision data and
simulation~\cite{JES2}.
For jets within the acceptance, the JES uncertainty varies
in the range 4--8\%
as a function of jet $\pT$ and $\eta$.  
Comparing MC and data the estimated systematic uncertainties are
  10\%\ and 1--2\%\ for the jet energy resolution (JER) and the efficiency,
respectively.
The uncertainty on the efficiency of the $b$-tagging algorithm 
has been estimated to be 6\% for $b$-quark jets, 
based on $b$-tagging calibration studies \cite{combNN}.

The uncertainty in the kinematic distributions of the $t \overline{t}$ signal events
gives rise to systematic uncertainties in
the signal acceptance, with contributions from the choice of generator, the modeling of initial- and final-state
radiation (ISR/FSR) and the choice of the PDF set.
The generator uncertainty is evaluated by comparing the  {\sc MC@NLO} predictions  with those of
POWHEG~\cite{powheg,Frixione:2007vw,Alioli:2010xd} interfaced to
either HERWIG or PYTHIA.
The uncertainty due to ISR/FSR is evaluated using the {\sc AcerMC} generator~\cite{Acer}
interfaced to the PYTHIA shower model, and by varying the parameters controlling ISR
and FSR in a range consistent with experimental data~\cite{Atlas_sim}.
Finally, the PDF uncertainty is evaluated using a range of
current PDF sets~\cite{PDF,MSTW,CTEQ5}.
The dominant uncertainty in this category of systematic uncertainties is the modelling of ISR/FSR.

The $\tau$ ID uncertainty is derived from a template fit to a $Z \rightarrow \tau^{+} \tau^{-}$ 
data sample selected with the same $\mu$ and $\tau$ 
candidate requirements as the sample for this analysis, but with fewer than two jets and 
$m_{\text T}<20$ GeV to remove $W+$jets events. The fit relies on the $W+1$ jet data
sample for a 
background template and $Z\rightarrow \tau^{+} \tau^{-}$ MC events for a signal template.
The uncertainty includes the statistical uncertainty of the data samples, the uncertainty
in the $\Zg$ cross section measured by ATLAS ~\cite{ZXsec} (excluding luminosity uncertainty)  and jet energy
scale uncertainty. It also includes
the uncertainty on the number of misidentified electrons ($<0.5$\%, determined from $Z\rightarrow e^{+}e^{-}$ data).

\begin{table}[!hbt]
\begin{center}
\caption{\label{tab:fits-systematics} Relative systematic uncertainties, in \%, for the cross-section measurement.
The first column gives the source of systematic uncertainty, ID/Trigger stands 
for the combined uncertainty 
of lepton identification and lepton trigger.
  The $\tau$ ID uncertainty includes electrons misidentified as $\tau$ leptons. The second and third columms give the
channel.}
\begin{tabular}{l|cc} \hline \hline
Source& $\mu+\tau$ & $e+\tau$\\ \hline 
$\mu$ (ID/Trigger) & $-$1.1 $/+$1.5 & -- \\ \hline
$e$ (ID/Trigger) & -- & $\pm$2.9  \\ \hline
JES & $-$2.0/$+$2.2 & $-$1.9 $/+$2.8 \\
JER & $\pm$1.0 & $\pm$1.2 \\
ISR/FSR         & $\pm$4.8 & $\pm$3.5\\
Generator & $\pm$0.7 & $\pm$0.7 \\     
PDF & $\pm$2.0 & $\pm$2.1 \\ 
$b$-tag & $-$7.7/$+$9.0 & $-$7.5/$+$8.9 \\ \hline
$\tau_1$ ID & $-$3.0/$+$3.2 & $-$2.7/$+$3.0 \\
$\tau_3$ ID & $-$3.1/$+$3.4 & $-$2.9/$+$3.2  \\ \hline
\hline
\end{tabular}

\end{center}
\end{table}

The effect of these variations on the final result is 
evaluated by varying each source of systematic uncertainty by $\pm 1\sigma$, applying
the selection cuts and recalculating the cross section.

The uncertainties obtained for the fit method using the 0 $b$-tag background template 
are shown in Table~\ref{tab:fits-systematics}.
The systematic uncertainties for the matrix method are very similar.
The uncertainty on the measured integrated luminosity is 3.7\% \cite{lumi}.

\section{Measuring the $t \overline{t}$ Cross Section}
\label{sec:crosssection}

The cross section is derived  from the number of observed OS-SS signal
events in the $\geq 1$ $b$-tag data sample assuming the only top quark decay mode is $t\rightarrow Wb$, and
subtracting from that number the small
contribution from $t \overline{t} \rightarrow e+\ell$ (from
electrons faking $\tau$ leptons)  
and $\tau$ leptons from $Z\rightarrow \tau^{+} \tau^{-}$ 
(Table~\ref{table:SelectedEvents}).
The systematic uncertainties are estimated as the quadratic sum of all uncertainties  
given in Table
\ref{tab:fits-systematics}, which includes the uncertainty from the subtraction.

The results are given separately for $\tau_1$ and $\tau_3$
and then combined (weighted by their statistical uncertainty
and assuming all systematic uncertainties other than from $\tau$ ID
are fully correlated). The results using the 0 $b$-tag background template 
are shown in Table  \ref{tab:fit_Xsec}.

\begin{table}[t]
\begin{center}
\caption{\label{tab:fit_Xsec} Measured cross section from the $\tau_1$ and $\tau_3$ samples, 
as well as the combination ($\tau_1$+$\tau_3$) for each channel separately. The uncertainty in the integrated 
luminosity (3.7\%) is not included.}
{\renewcommand\arraystretch{1.25}
\begin{tabular}{c|c}
\hline
\hline
& $\mu+\tau$ \\
\hline
$\tau_1$  &$189\pm 16 \stat \pm20 \syst$ pb \\
$\tau_3$ &$180\pm 40 \stat \pm 21 \syst $ pb \\ \hline
$\tau_1$+$\tau_3$ &$186\pm 15 \stat \pm 20\syst $ pb \\ \hline \hline
& $e+\tau$ \\ \hline
$\tau_1$  &$190\pm 20 \stat \pm 20\syst $ pb \\
$\tau_3$  &$170\pm 50 \stat \pm 21 \syst $ pb \\ \hline 
$\tau_1$+$\tau_3$ & $187\pm 18 \stat \pm 20\syst $ pb \\
\hline\hline
\end{tabular}
}
\end{center}
\end{table}

The results for the $\mu+\tau$ and $e+\tau$ channels are combined taking into 
account the correlated
uncertainties using the BLUE (Best Linear Unbiased Estimator) technique \cite{BLUE}.
Combining them does not improve
the systematic uncertainty as the systematic uncertainties are almost 100\% correlated.

The results for each lepton type are:
\begin{equation*}
\mu+\tau:~~\sigma_{t\overline{t}} = 186\pm 15  \stat \pm 20 \syst \pm 7~\text{(lumi.)}~ \ \ {\rm pb},
\end{equation*}
\begin{equation*}
e+\tau:~~\sigma_{t\overline{t}} = 187\pm 18  \stat \pm 20 \syst \pm 7~\text{(lumi.)}~ \ \ {\rm pb},
\end{equation*}

Combining both channels one obtains:
\begin{equation*}
\sigma_{t\overline{t}} = 186\pm 13  \stat \pm20 \syst \pm 7~\text{(lumi.)}~ \ \ {\rm pb}
\end{equation*}

To check the fit measurements, the cross sections can be calculated  
using the matrix method and the results obtained with the $W+1$ jet background to minimize
the correlation with the fit results. The combination of the matrix method and the fit results 
with the BLUE method
show  they are compatible at the 45\% and 10\% confidence level for $\mu+\tau$ and $e+\tau$, respectively.

\section{Conclusions}

The cross section for $t \overline{t}$ production in $pp$ collisions at 7
TeV has been measured in the $\mu + \tau$ and the $e+\tau$
channels in which the $\tau$
decays hadronically. The number of $\tau$ leptons in these channels
has been extracted using multivariate discriminators to separate
$\tau$ leptons from electrons and jets misidentified as hadronically decaying $\tau$ leptons. 
These numbers were obtained
by fitting the discriminator outputs and checked with a matrix method.
Combining the measurements
from $\mu+\tau$ and $e+\tau$ events, 
the cross section is measured to be 
\begin{equation*}
\sigma_{t\overline{t}} = 186\pm 13  \stat \pm20 \syst \pm 7~\text{(lumi.)}~ \ \ {\rm pb},
\end{equation*}
in good agreement with the cross section measured by ATLAS in other channels ~\cite{{ATLAS_ljets},{ATLAS700_dil}}, with the cross-section measurement by the CMS Collaboration~\cite{CMSltau,CMSljets} and 
with the SM prediction,  $164^{+11}_{-16}$~pb ~\cite{HATHOR}.


\section{Acknowledgements}

We thank CERN for the very successful operation of the LHC, as well as the
support staff from our institutions without whom ATLAS could not be
operated efficiently.

We acknowledge the support of ANPCyT, Argentina; YerPhI, Armenia; ARC,
Australia; BMWF, Austria; ANAS, Azerbaijan; SSTC, Belarus; CNPq and FAPESP,
Brazil; NSERC, NRC and CFI, Canada; CERN; CONICYT, Chile; CAS, MOST and NSFC,
China; COLCIENCIAS, Colombia; MSMT CR, MPO CR and VSC CR, Czech Republic;
DNRF, DNSRC and Lundbeck Foundation, Denmark; EPLANET and ERC, European Union;
IN2P3-CNRS, CEA-DSM/IRFU, France; GNAS, Georgia; BMBF, DFG, HGF, MPG and AvH
Foundation, Germany; GSRT, Greece; ISF, MINERVA, GIF, DIP and Benoziyo Center,
Israel; INFN, Italy; MEXT and JSPS, Japan; CNRST, Morocco; FOM and NWO,
Netherlands; RCN, Norway; MNiSW, Poland; GRICES and FCT, Portugal; MERYS
(MECTS), Romania; MES of Russia and ROSATOM, Russian Federation; JINR; MSTD,
Serbia; MSSR, Slovakia; ARRS and MVZT, Slovenia; DST/NRF, South Africa;
MICINN, Spain; SRC and Wallenberg Foundation, Sweden; SER, SNSF and Cantons of
Bern and Geneva, Switzerland; NSC, Taiwan; TAEK, Turkey; STFC, the Royal
Society and Leverhulme Trust, United Kingdom; DOE and NSF, United States of
America.

The crucial computing support from all WLCG partners is acknowledged
gratefully, in particular from CERN and the ATLAS Tier-1 facilities at
TRIUMF (Canada), NDGF (Denmark, Norway, Sweden), CC-IN2P3 (France),
KIT/GridKA (Germany), INFN-CNAF (Italy), NL-T1 (Netherlands), PIC (Spain),
ASGC (Taiwan), RAL (UK) and BNL (USA) and in the Tier-2 facilities
worldwide.

\clearpage
\onecolumn
\begin{flushleft}
{\Large The ATLAS Collaboration}

\bigskip

G.~Aad$^{\rm 48}$,
B.~Abbott$^{\rm 111}$,
J.~Abdallah$^{\rm 11}$,
S.~Abdel~Khalek$^{\rm 115}$,
A.A.~Abdelalim$^{\rm 49}$,
O.~Abdinov$^{\rm 10}$,
B.~Abi$^{\rm 112}$,
M.~Abolins$^{\rm 88}$,
O.S.~AbouZeid$^{\rm 158}$,
H.~Abramowicz$^{\rm 153}$
% ...
\bigskip

% ...
$^{10}$ Institute of Physics, Azerbaijan Academy of Sciences, Baku, Azerbaijan\\
$^{11}$ Institut de F\'isica d'Altes Energies and Departament de F\'isica de la Universitat Aut\`onoma  de Barcelona and ICREA, Barcelona, Spain\\
% ...
$^{48}$ Fakult\"{a}t f\"{u}r Mathematik und Physik, Albert-Ludwigs-Universit\"{a}t, Freiburg i.Br., Germany\\
$^{49}$ Section de Physique, Universit\'e de Gen\`eve, Geneva, Switzerland\\
% ...
$^{88}$ Department of Physics and Astronomy, Michigan State University, East Lansing MI, United States of America\\
% ...
$^{111}$ Homer L. Dodge Department of Physics and Astronomy, University of Oklahoma, Norman OK, United States of America\\
$^{112}$ Department of Physics, Oklahoma State University, Stillwater OK, United States of America\\
% ...
$^{115}$ LAL, Univ. Paris-Sud and CNRS/IN2P3, Orsay, France\\
% ...
$^{153}$ Raymond and Beverly Sackler School of Physics and Astronomy, Tel Aviv University, Tel Aviv, Israel\\
% ...
$^{158}$ Department of Physics, University of Toronto, Toronto ON, Canada\\
% ...
\end{flushleft}

\end{document}